\documentclass[11pt]{article}
\usepackage{moriond,epsfig}

\def\be{\begin{equation}}
\def\ee{\end{equation}}
\def\bea{\begin{eqnarray}}
\def\eea{\end{eqnarray}}

\begin{document}
\vspace*{4cm}
\title{SINGLE W AND Z PRODUCTION, ASYMMETRIES, AND V+JETS AT THE
TEVATRON}

\author{ HANG YIN}

\address{(On behalf of CDF and D0 Collaborations)\\ 
Fermi National Accelerator Laboratory, \\
Pine Street and Kirk Road, Batavia, IL, 60510 US}

\maketitle\abstracts{We present the most recent $W$ and $Z$ production, asymmetries, and
$V$+jets results
from the Tevatron collider at $\sqrt{s}$ = 1.96 TeV, analyzing data
collected by
CDF and D0 detectors~\cite{cdf_detector,d0_detector} between 2002-2011. The results include the
measurements
of the $W$ and $Z$ bosons properties, include boson $p_T$ and
asymmetries measurements,
and also the $W$ and $Z$ boson plus jets productions.
Those measurements provide precision tests on the electroweak theory in
Standard Model (SM),
parton distribution functions (PDFs)~\cite{pdf}, and high order theoretical
predictions.}

\section{Introduction}
The Tevatron has been shutdown at 2011, after that, there are plenty of analyses related
to the single $W$ and $Z$ boson have been performed. With precision 
measurement of $W$/$Z$ properties and asymmetries, several
critical tests have been done for electroweak theory in the SM. 
The Tevatron is a $p\bar{p}$ collider, the $W$
and $Z$ are produced with the $valence$ $quark$, and in the same time, at the LHC, in the
productions of single $W$ and $Z$ boson, the contributions from $sea$ $quark$
and $gluon$ become larger. Thus, single $W$ and $Z$ measurements
at the Tevatron are complementary for LHC. Furthermore, with knowing incoming $quark$
direction, the $Z$ forward-backward charge asymmetry at the Tevatron is more sensitive
compared with the LHC.

There are many of $W/Z+$ jets measurements in the past year, with
CDF/D0 full data-set, the statistics uncertainties has been highly
suppressed. All of those measurements provides a precision test on the
perturbative quantum chromodynamic (QCD) formalism, also the quark PDFs.

In this note, we will first review two single $Z$ related analyses, which are
weak mixing angle measurement~\cite{cdf_stw} and $Z$ $p_T$ measurements~\cite{cdf_zpt}. 
After that, there will be short review on the 
Tevatron $W/Z+$ jets measurements.

\section{Single $W$ and $Z$ productions}
\subsection{Weak mixing angle}
Weak mixing angle is one of fundamental parameters in SM. It is a key parameter
related to the electroweak couples, for both charge current ($W$)
and neutrino current ($Z$). The weak mixing angle is a running
parameters in a wide region of central of energy, there are many measurements
have been done before, like atomic parity violation~\cite{APV}, Milloer scattering~\cite{milloer}
, and NuTeV~\cite{nutev}. In the $Z$ peak
region, the most precision measurements are from LEP $b$ quark asymmetries,
and SLD Left-Right hand asymmetries ($A_{LR}$)~\cite{lep}. The results from 
those two measurements are deviated by three standard deviations in different directions.
Recently at the hadronic collider, CDF~\cite{cdf_stw_old}, D0~\cite{dzero_1fb,dzero_5fb}, and CMS~\cite{cms_stw} have been performed this measurement,
which show reasonable agreement with world average value. 
Due to the limitations from the parton distribution functions (PDFs)
and quark fragments, the dominated systematic uncertainty comes from
PDFs, which can be possibly suppressed by the update of PDFs sets.

In the SM, with boosted effects from the QCD radiation, the Collin-Soper (CS) frame~\cite{cs_frame}
is employed to reduce the impact from the boson $p_T$. And in the CS frame,
the general expression for the angular distribution is described by the 
polar ($\theta$) and azimuthal ($\phi$) angles of the decay-electrons,
the $Z$ boson differential cross section~\cite{z_diffx} can be wrote as:
\begin{eqnarray}
\frac{dN}{d\Omega} = &&(1+cos^2\theta) + A_0\frac{1}{2}(1-3cos^2\theta) + \nonumber \\
&&A_1 sin2\theta cos\phi + A_2\frac{1}{2}sin^2\theta cos 2\phi + \nonumber\\
&&A_3 sin\theta cos\phi + A_4 cos\theta + A_5 sin^2\theta sin2\phi + \nonumber\\
&&A_5 sin^2\theta sin 2\phi + A_6 sin 2\theta sin \phi + A_7 sin\theta sin\phi
\end{eqnarray}
where, the $A_{0}$ and $A_{4}$ are extracted from $\cos \theta$ distribution, 
and $A_2$ and $A_3$ are extracted from $\phi$ distribution, while
$A_5$ to $A_7$ are expected to be zero.
The $A_4$ term is related to the $Z/\gamma^*$ forward and backward charge
asymmetry distribution. There are two sources contribute to the asymmetry,
one is the photon vector and $Z-$axial interference, 
this asymmetric component is proportional $g_A$,
another one is the $Z-$boson amplitude self interference, 
which has a coupling factor that is a product of $g_V/g_A$ 
from the electron and quark vertices, related to the $\sin^2\theta_W$.
At the Born level, this product is 
\begin{eqnarray*}
(1-4|Q_e|\sin^2\theta_W)(1-4|Q_q|\sin^2\theta_W)
\end{eqnarray*}
where $e$ and $q$ represent the electron and quark, $q$
is the charge of a light quark ($u$, $d$, or $s$).

With 2.1 fb$^{-1}$ data, CDF uses the same selection cuts and tuned Monte Carlo (MC)
as $Z$ angular coefficients~\cite{cdf_angu}. A new method has been developed to measure weak mixing angle. 
Two inputs used to extract $\sin^2\theta_W$, a measured value of $A_4$
and $A_4$ templates with different $\sin^2\theta_W$ input. The measured 
value, denoted as $\bar{A_4}$, is a cross section weighted average in 
the dielectron mass region between 66 - 116 GeV/$c^2$.
The templates sets are calculated with same dielectron kinematic region,
which can provide the relationship between input $\sin^2\theta_W$
and $\bar{A_4}$. The QCD calculation of $A_4$ include a LEP-like
implementation of electroweak radiative corrections.

The $\bar{A_4}$ is derived from the previous measurement of electron
decay angular distribution coefficients. The coefficients are measured
in bins of the dielectron pair $p_T$, it is a data driven measurement,
by comparing the observed $\cos\theta$ distribution against the simulated
templates with different $A_4$ input value. The coefficients are adjusted
in bins of the $ee-$pair $p_T$.
The simulation of Drell-Yan pair production uses the {\sc PYTHIA}~\cite{pythia} generator,
combined with CDF detector simulation programs. With CTEQ5L nucleon parton
distribution functions (PDFs), and {\sc PHOTOS}~\cite{photos} to provide a good model
of QED final state radiation, the simulation provide a proper MC sample,
which can well describe the data.

As shown in Figure~\ref{fig:cdf_stw_1}, the $\cos\theta$ distribution
for combined central-central (CC) and central-plus (CP) topology dielectron.
For the separated CC and CP
topology $\cos\theta$ distribution, the $\chi^2/dof$'s between the simulation
and the data are 49.0/50 and 46.9/46. 
The data $A_4$ values are measured in five $p_T$ bins: 0-10, 10-20, 20-35,
35-55 and $>$ 55 GeV/$c^2$. Since both $A_0$ and $A_4$ can change 
the $\cos\theta$ distribution, a simultaneously fitting of both $A_0$ and
$A_4$ is performed. The physics model $\cos\theta$ distribution is varied
using an event reweighting method. The event weight is defined as:

\begin{eqnarray*}
 \omega = \frac{N(\theta, A_0^{'}, A_4^{'})}{N(\theta, A_0, A_4)}
\end{eqnarray*}
where $A_0^{'}$ and $A_4^{'}$ denote to the variations of those two parameters,
at the same time, $A_0$ and $A_4$ are represent to the base physics model
angular coefficients. The measured values of $A_4(p_T)$ is incorporated into
the physics model for the calculation. The calculation gives
\begin{eqnarray*}
\bar{A_4} = 0.1100\pm0.0079(stat)\pm0.0003 (syst)
\end{eqnarray*}
where the first uncertainty is the statistical uncertainty and
the second the systematic uncertainty. The standard model predictions
of $\bar{A_4}$ of various input values of $M_W$ (or $\sin^2\theta_W$)
calculated by {\sc RESBOS}~\cite{resbos} are shown in Figure~\ref{fig:cdf_stw_1}. 
The indirect measurement of $M_W$:
\begin{eqnarray*}
 M_W(indirect) = 80.297\pm 0.055 GeV/c^2
\end{eqnarray*}
where the uncertainty includes both measurement and prediction uncertainties.
The other $W$-mass measurements shown in Figure~\ref{fig:cdf_stw_1}
are from combinations of Tevatron and LEP/SLD measurements~\cite{wmass_com}:
\begin{eqnarray*}
 M_W &=& 80.385\pm 0.015 GeV/c^2,~direct \\
     &=& 80.365\pm0.020 GeV/C^2,~Z~pole,
\end{eqnarray*}
where the direct is the combination of LEP2 and Tevatron $W$ mass measurements,
and $Z$ pole is an indirect measurement from electroweak SM fits to LEP1/SLD
$Z$-pole measurements with the top mass measurements.
With $\bar{A_4}$, the extracted $\sin^2\theta^{lept}_{eff}$ is
\begin{eqnarray*}
\sin^2\theta^{lept}_{eff} = 0.2328 \pm 0.0010
\end{eqnarray*}
which is consistent with the Tevatron value from D0~\cite{dzero_5fb}.

\begingroup
\begin{figure}[htbp]
\epsfig{file=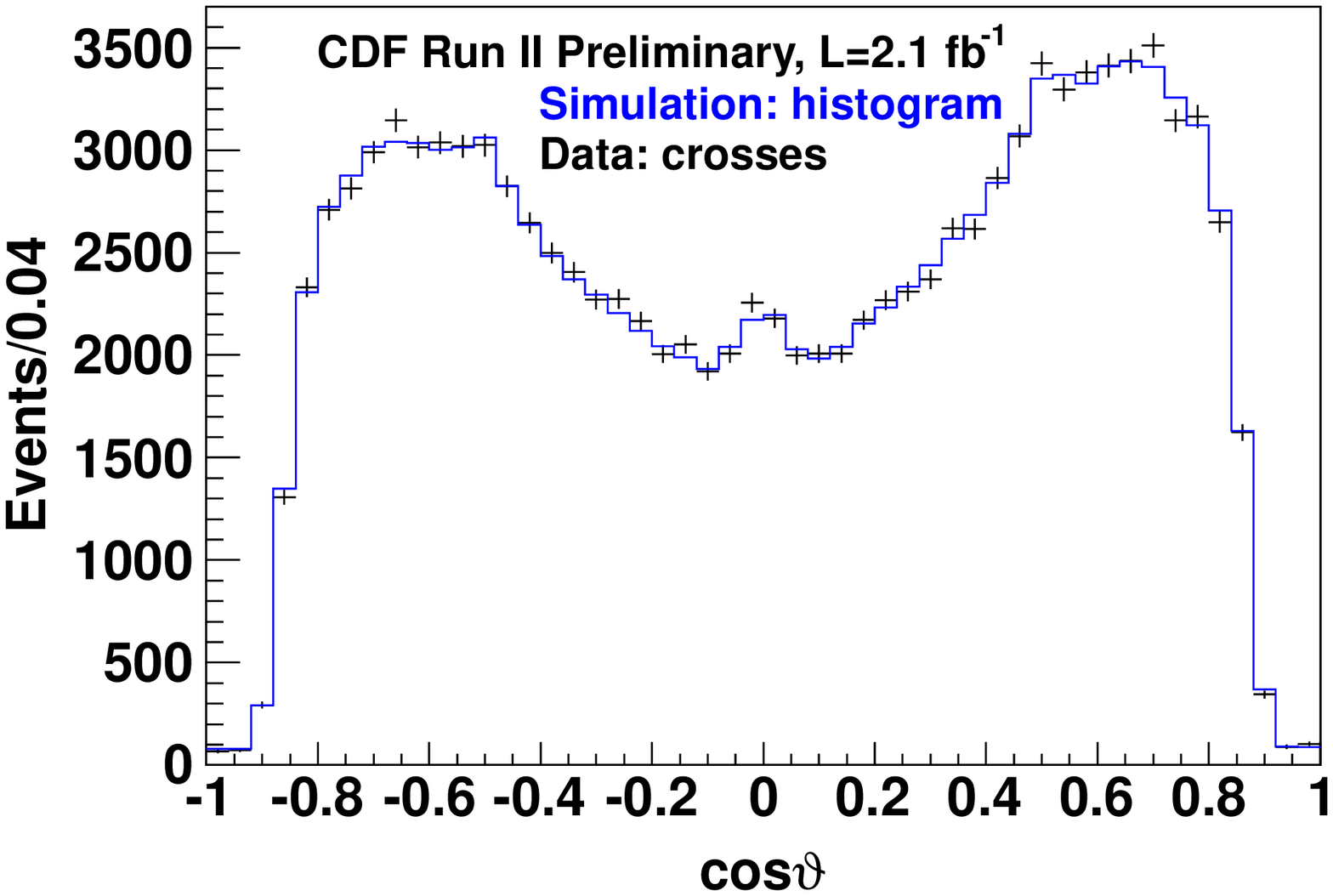,width=250pt,height=175pt}
\epsfig{file=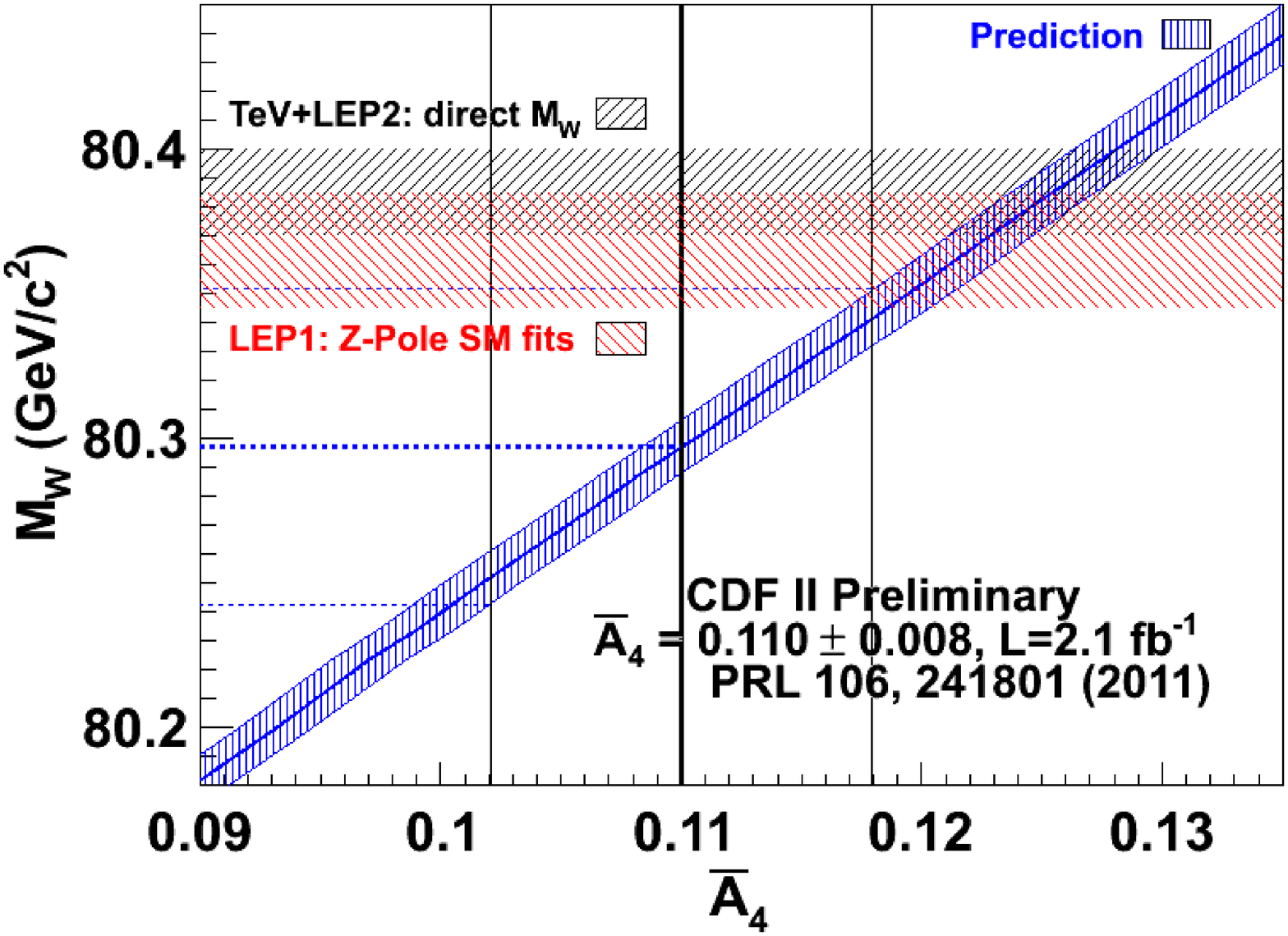,width=250pt,height=175pt}
\caption{\small The left plot is $\cos\theta$ distribution for combined CC and CP
$ee-$pair topologies. The black crosses are data after background subtraction,
the solid (blue) line is the simulation.
The right plot is standard model predictions of $\bar{A_4}$ in relation to
the input $M_W$ calculated by {\sc RESBOS}: The prediction is the solid (blue)
diagonal line and its one standard deviation uncertainty interval is the band.
The $\bar{A_4}$ measurement is the bold vertical line, and its one standard
deviation limits are the lighter vertical lines. The hatched horizontal bands
are uncertainty limits from other $W$ mass measurements.}
\label{fig:cdf_stw_1}
\end{figure}
\endgroup

\subsection{$Z$ Transverse momentum cross section measurement}
Initial state QCD radiation from the colliding
parton can change the kinematic of the Drell-Yan process~\cite{drellyan} system, 
give $Z$ boson a transverse momentum, thus the precision measurement
of $Z$ boson $p_T$ can provide a stringent test on the higher order
QCD perturbative calculation.

With 2.1 fb$^{-1}$ of integrated luminosity data, CDF performed a precision
measurement of transverse momentum cross section of $e^+e^-$ pairs in the
$Z-$boson mass region of 66-116 GeV/$c^2$. In order to get agreement between
data $p_T$ distribution and simulated $p_T$ distribution, the generator
level $p_T$ distribution is adjusted, bin-by-bin. The method uses the
data-to-simulation ratio of the number of reconstructed events in $p_T$
bins as an iterative adjustment estimator for the generator level $p_T$ bins.
Figure~\ref{fig:cdf_zpt_1} is the generator level $p_T$ correction function
that makes the data-to-simulation ratio uniform.

\begingroup
\begin{figure}[htbp]
\epsfig{file=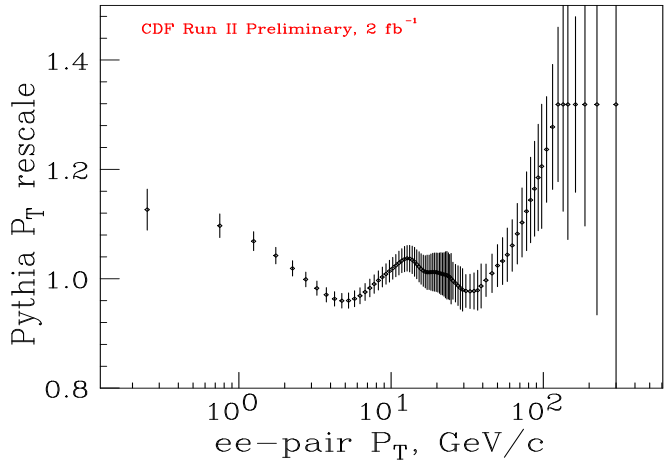,width=250pt,height=175pt}
\caption{\small The correction function applied to the generator level
$\Delta N/\Delta p_T$ distribution that makes flat the ratio of observed
data to the simulated data as function of boson $p_T$. The points are at
the central of the $p_T$ bins.}
\label{fig:cdf_zpt_1}
\end{figure}
\endgroup

In Figure~\ref{fig:cdf_zpt_2}, the measured cross section is compared 
with two quantum chromodynamic calculations. 
One is {\sc FEWZ2}~\cite{fewz2}, a fixed-order perturbative calculation at $\O(\alpha_s^2)$,
and the other is {\sc RESBOS}, which combines perturbative predictions at high transverse momentum
with the gluon resummation formalism at low transverse momentum. Comparisons
of the measurement with calculations show reasonable agreement.
The ratio of the measured cross section to the {\sc RESBOS} prediction in the
low $p_T$ region is shown in Figure~\ref{fig:cdf_zpt_2}.

\begingroup
\begin{figure}[]
\epsfig{file=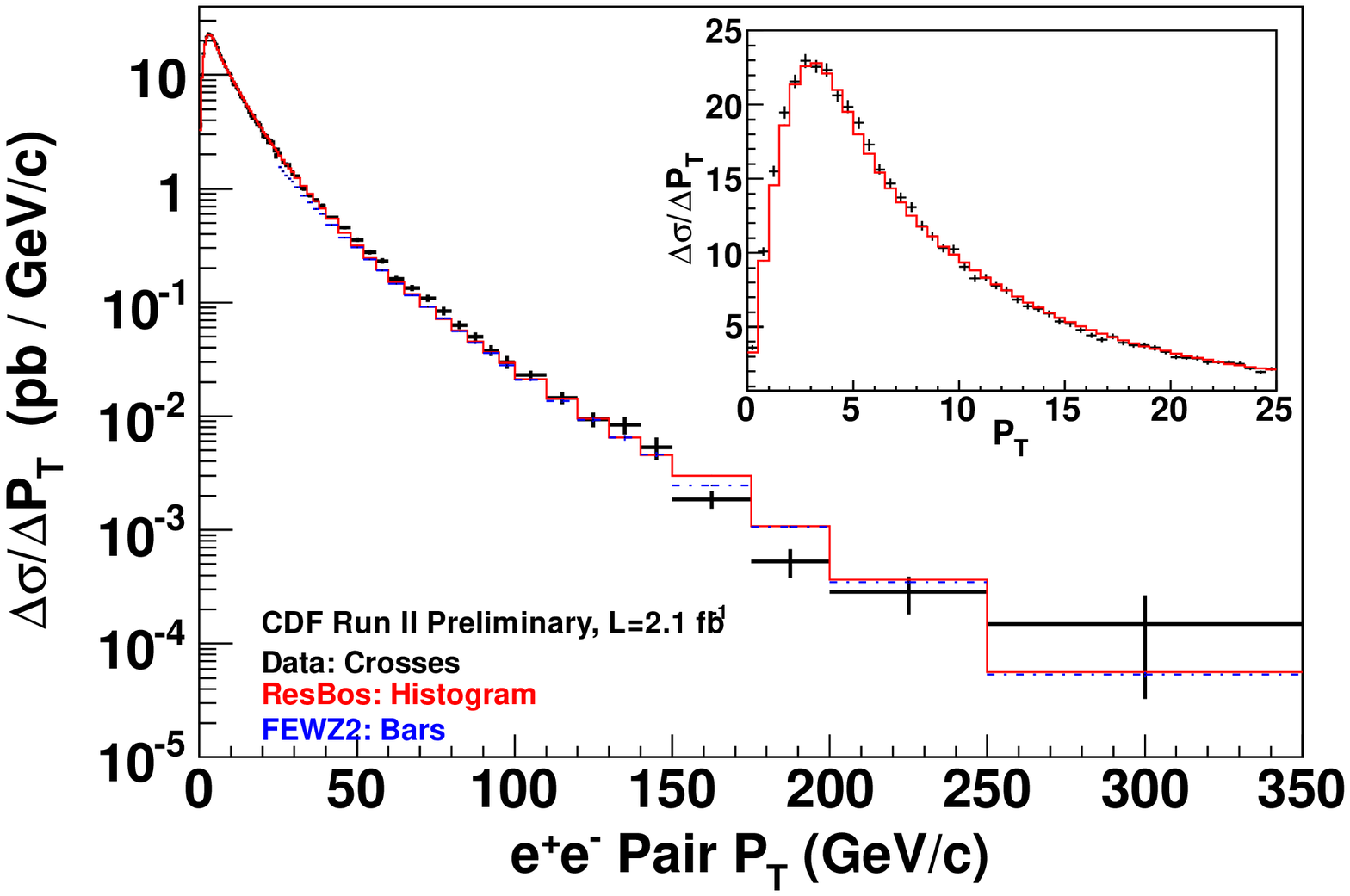,width=250pt,height=175pt}
\epsfig{file=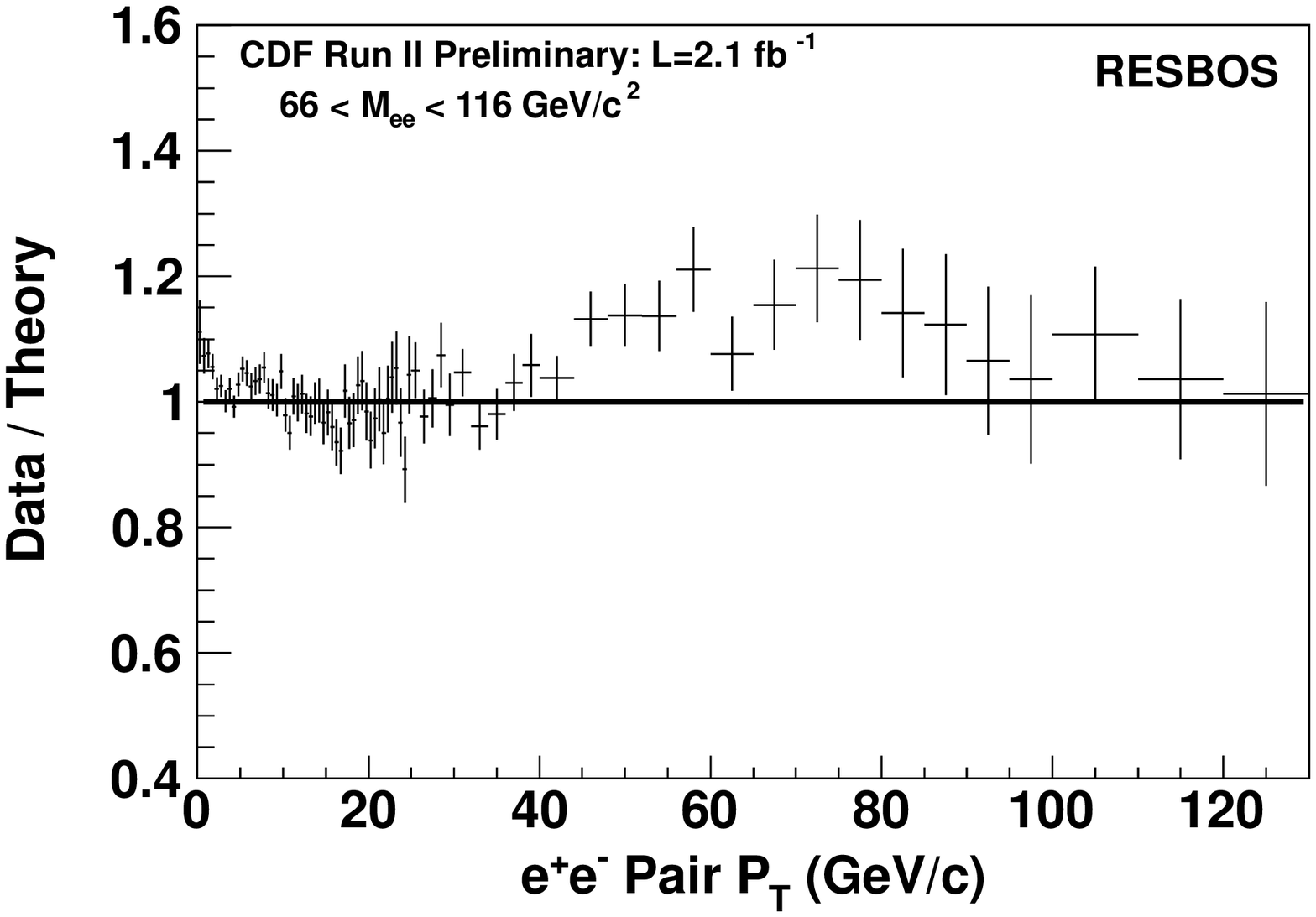,width=250pt,height=175pt}
\caption{\small The left plot is $\Delta \sigma/\Delta p_T$ cross section
versus $Z$ boson $p_T$. The solid (black) crosses are the data and all
uncertainties except the integrated luminosity uncertainty are combined
and plotted. The solid (red) histogram is the {\sc RESBOS} calculation.
The dash-dotted (blue) bars of the $p_T > $ 25 GeV/$c$ region are the 
{\sc FEWZ2} calculation. The inset is the $p_T < $ 25 GeV/$c$ region
with a linear ordinate scale.
The right plot is the ratio of the measured cross section to the {\sc RESBOS}
prediction in the $p_T < $ 130 GeV/$c$ region. The {\sc RESBOS} total
cross section is normalized to the data.}
\label{fig:cdf_zpt_2}
\end{figure}
\endgroup

%%%%%%%%%%%%%%%%%%%%%%%%%%%%
%%% V+jets results
%%%%%%%%%%%%%%%%%%%%%%%%%%%%
\section{$V$ + Jets measurements}
$W$, $Z$ plus jets have been studied in both CDF and D0. Those processes are
very important backgrounds for new physics search, and provide stringent
test on the perturbative QCD calculation, and the theoretical predictions
are suffered from large uncertainties. For some special channel, the quark
PDFs can be constraint, like $W+c$.

\subsection{$W$+Jets measurement}
Using a 3.7 fb$^{-1}$ dataset collected by the D0 detector, D0
measured inclusive $W(\rightarrow e\nu)+n-jets~(n\geq~1,2,3,4)$
production~\cite{dzero_wjet}. Differential cross section are presented as a function
of the jet rapidity ($y$), lepton transverse momentum ($p_T$) 
and pseudo-rapidity ($\eta$), the scalar sum of the transverse energies of
the $W$ and all jets ($H_T$), leading dijet $p_T$ and mass, dijet rapidity,
opening angle, azimuthal angular. And compared the measured cross
section with plenty of MC predictions.
% In Figure~\ref{fig:dzero_wjet} 
%shows measured jet rapidity compared with various theortical predictions.

%\begingroup
%\begin{figure}[htbp]
%\epsfig{file=plots/DZero_W_Jets.eps,width=175pt,height=250pt}
%\caption{\small Measurement of the $n^{jet}-jet$ rapidity distributions
%in inclusive $W+n-$jet events for n = 1-4 and comparison to varous 
%theoretical predictions. Lower panes show theory/data comparisons for each
%of the $n-$jet multiplicity bin results separately.}
%\label{fig:dzero_wjet}
%\end{figure}
%\endgroup

\subsection{$W+b$ measurement}

A measurement of $W+b$ production cross section~\cite{cdf_wb} with up to two jets
has been published by CDF collaboration, the measured cross section reported
is $\sigma \cdot \mathcal{B}(W\rightarrow \mathit{l}\nu) = 2.74\pm0.27(stat.)\pm0.42(syst.)$ pb ($\mathit{l} = e, \mu$), with ~3 standard deviation from the predictions.
With 6.1 fb$^{-1}$ data, D0 measured an inclusive cross section~\cite{dzero_wb} of
$\sigma (W(\rightarrow \mathit{l}\nu)+b+X)= 1.05\pm0.12(stat+syst.)$ pb
for $|\eta^{\mathit{l}}|<$ 1.7. The results are in agreement with prediction from
next-to-leading order QCD calculation using {\sc MCFM}~\cite{mcfm}, and also
with predictions from the {\sc SHERPA}~\cite{sherpa} and {\sc MADGRAPH}~\cite{madgraph} MC event generators.

%\begingroup
%\begin{figure}[htbp]
%\epsfig{file=plots/DZero_Wb_msl.eps,width=250pt,height=175pt}
%\epsfig{file=plots/DZero_Wb_mjl.eps,width=250pt,height=175pt}
%\caption{\small The $\cos\theta$ distribution for combined CC and CP
%$ee-$pair topologies. The black crosses are data after background strubtraction,
%the solid (blue) line is the simulation.}
%\label{fig:dzero_wb}
%\end{figure}
%\endgroup

\subsection{$W+c$ measurement}
The associated production of the $W$ boson with a single charm quark in 
proton-antiproton collisions is described at the lowest order in the
standard model (SM) by quark-gluon fusion ($gq\rightarrow Wc$), where
$q$ denotes a $d$, $s$, or $b$ quark. At the Tevatron, the larger
$d$ quark PDF in the proton is compensated by the small quark-mixing
matrix element $|V_{cd}|$, only about 20\% of the total $Wc$ production is
comes from $gd\rightarrow Wc$, the majority contribution is comes from
strange quark-gluon fusion. Thus, the $Wc$ production cross section
is sensitive to the gluon and $s$ quark PDFs.

CDF collaboration report the first observation of the $Wc$ production~\cite{cdf_wc}, using
data corresponding to 4.3 fb$^{-1}$. Charm quark candidates are selected
through the identification of an electron or muon from charm-hadron
semileptonic decay within a hadronic jet, 
%the distributions of
%soft lepton $p_T$ are shown in Figure~\ref{fig:cdf_wc},
and a $Wc$ signal is observed
with a significance of 5.7 standard deviations. The production cross section
$\sigma Wc (p_Te>20 GeV/c, |\eta_e|<1.5)\times \mathcal{B}(W\rightarrow \mathit{l}\nu)$ is measured to be $13.6^{+3.4}_{-3.1}$ pb, in agreement with theoretical
expectations. The quark-mixing matrix element $V_{cs}$ is also derived,
$|V_{cs}|=1.08\pm0.16$ along with a lower limit of $|V_{cs}|>$ 0.71 at
95\% confidence level, asumming that the $Wc$ production throught $c$
to $s$ quark coupling is dominant.

%\begingroup
%\begin{figure}[htbp]
%\epsfig{file=plots/CDF_Wc.eps,width=175pt,height=250pt}
%\caption{\small The soft muon and soft electron $p_T$ distributions.}
%\label{fig:cdf_wc}
%\end{figure}
%\endgroup

\subsection{$Z+$ jets measurement}
Based on 6.0 fb$^{-1}$ data collected with CDF detector, CDF collaboration
reported a inclusive $Z/\gamma^*$ boson plus jets production~\cite{cdf_zjets}, the cross
sections are measured as a function of $p_T^{jet}$ and jet multiplicity for
jet in the region $p_T^{jet} \leq$ 2.1, results are compared with NLO
perturbative QCD predictions. One example of measured data compared
with different theoretical predictions is shown in Figure~\ref{fig:cdf_zjets}
for $Z/\gamma^*+\geq~1$ jet events.

\begingroup
\begin{figure}[htbp]
\epsfig{file=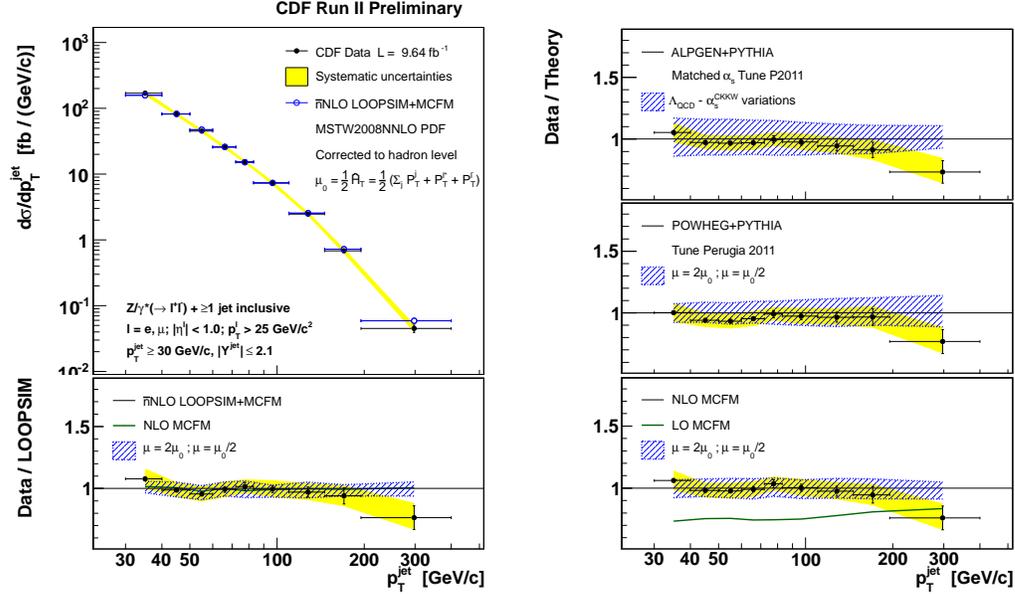,width=400pt,height=250pt}
\caption{\small Measured cross section in $Z/\gamma^*+\geq~1$ jet events as a function of inclusive $p_T^{jet}$. The left plot shows the comparison between 
measure data (blue point) compared with different theoretical predictions.
And the ratio between data and theory are shown in the right plots.}
\label{fig:cdf_zjets}
\end{figure}
\endgroup

\subsection{$Z+b$ measurements}
Studies of $Z$ boson production in association with jets from $b$ quarks,
or $b$ jets, provide important tests on the QCD predictions. And $Z+b$ is
also a major background for variety of physics processes. Both CDF and D0
measured the cross section~\cite{cdf_zb,dzero_zb} with RunII full data-set. The results are 
compared to the NLO predictions with various MC events generators.
As shown in Figure~\ref{fig:tev_zb}, the measured cross sections are
compared with different MC predictions for different various.

\begingroup
\begin{figure}[htbp]
\epsfig{file=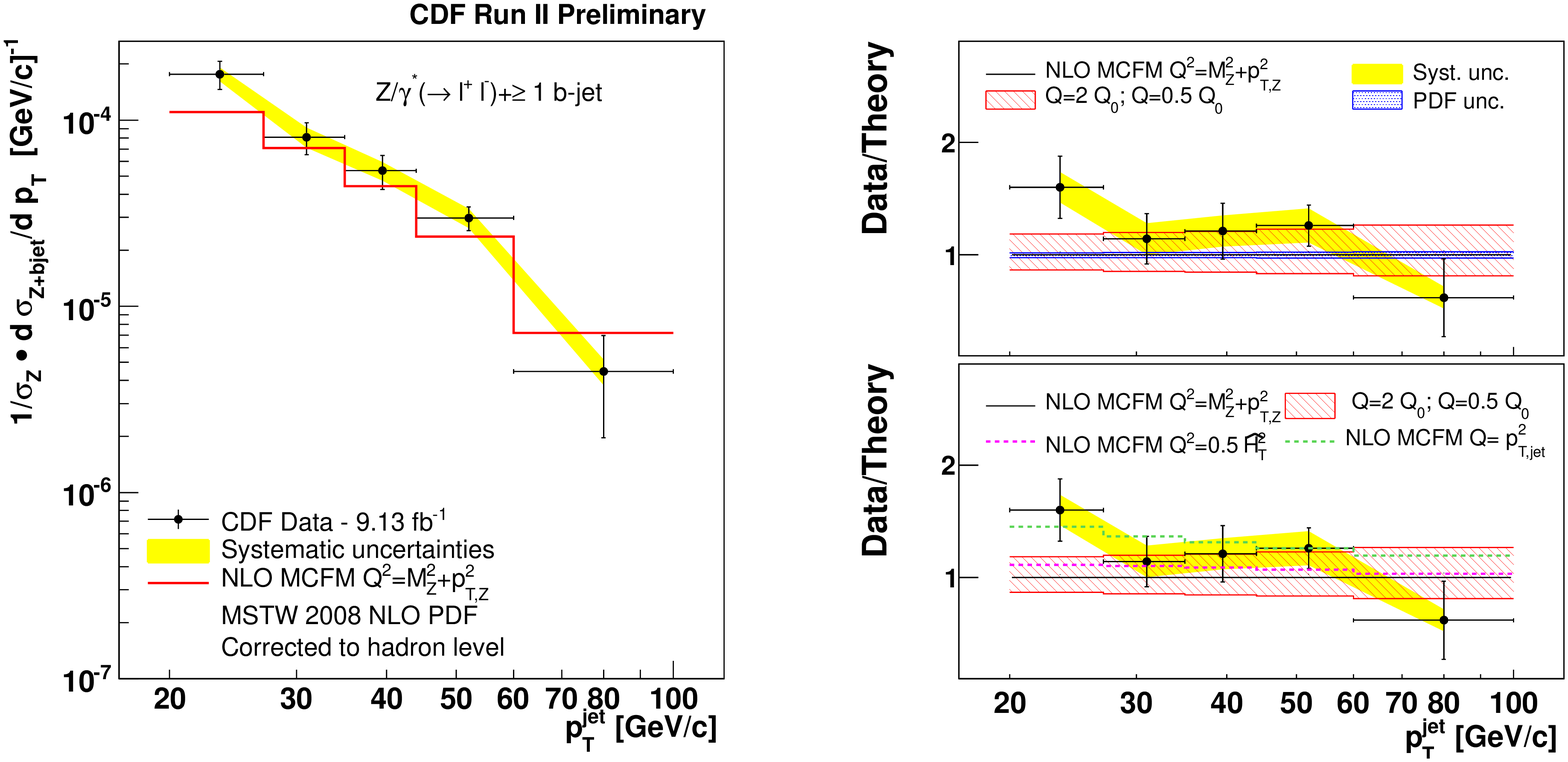,width=500pt,height=175pt}
\epsfig{file=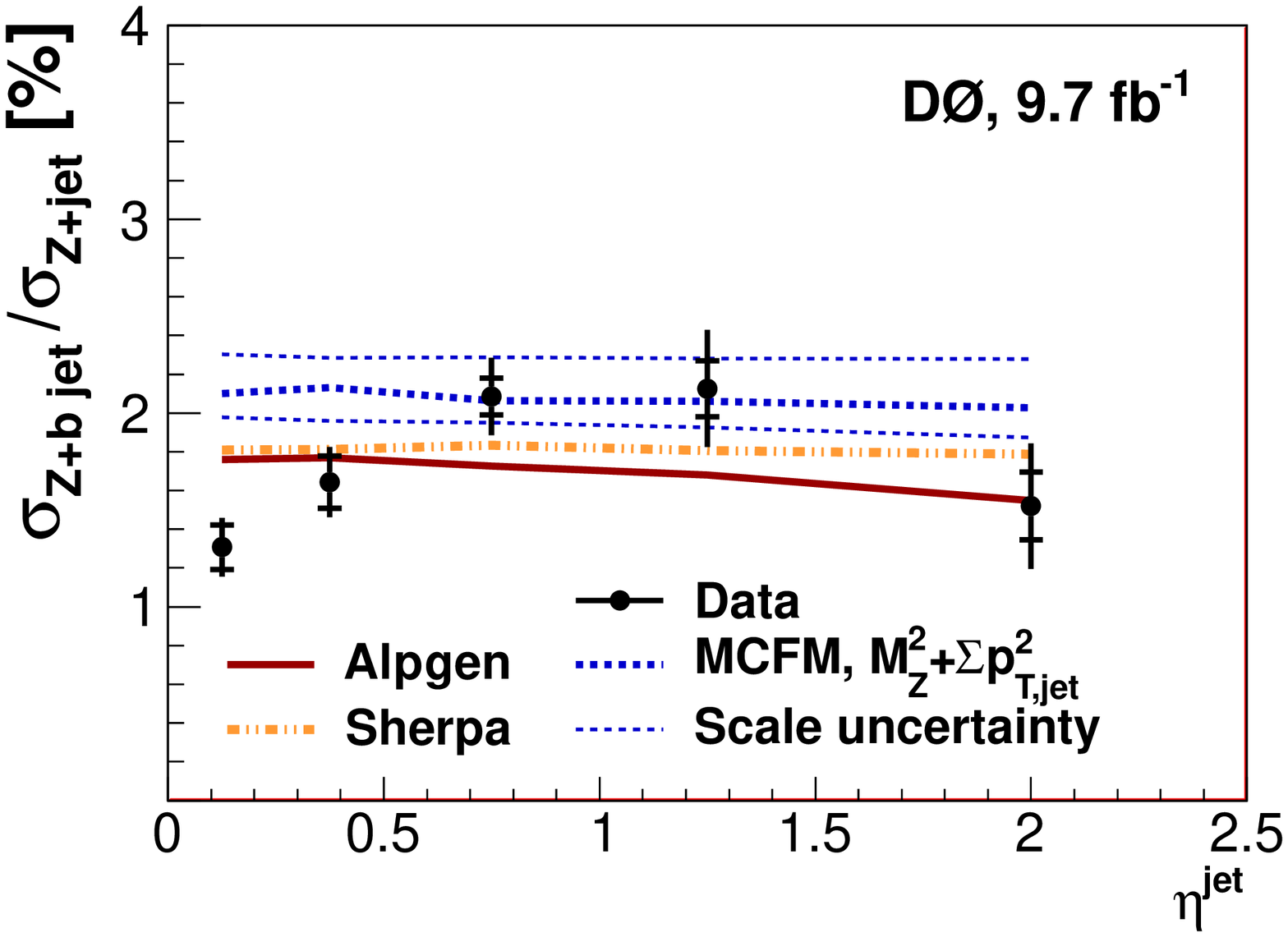,width=250pt,height=175pt}
\caption{\small The top-left plot shows the CDF measured differential
cross section as function of jet $p_T$, compared with MC predictions.
The top-right plot shows the ratio between data and theory.
The bottom plot shows D0 ratio of the differential cross
section as function of $\eta^{jet}$. 
The uncertainties on the data include statistical 
(inner error bar) and full uncertainties (entire error bar). The data
are compared to the prediction from {\sc ALPGEN}, {\sc SHERPA}, and
the {\sc MCFM} NLO calculation, where the band represents the variation
of the renormalization and factorization scale up and down by a factor
of two.}
\label{fig:tev_zb}
\end{figure}
\endgroup

\section{Summary}
CDF and D0 collaborations present plenty of precision measurements
of the single $W$ and $Z$ properties, $V+$jets. Those precision measurements
provide stringent tests on the electroweak theory in the SM, high order perturbative QCD calculation,
and quark PDFs. And with Tevatron full dataset, there will be more 
precision measurements coming soon, most of them will become to the legacy 
measurement at the Tevatron.


\section*{References}
\begin{thebibliography}{99}
\bibitem{cdf_detector} CDF II Detector Technical Design Report No. Fermilab-Pub-96/390-E; D. Acosta et al. (CDF Collaboration), Phys. Rev. D {\bf 71}, 052003 (2005).

\bibitem{d0_detector} V.M. Abazov, et al., (D0 Collaboration), Nucl. Instrum. Methods Phys. Res. A {\bf 565} (2006) 463.

\bibitem{pdf} J. Pumplin et al., J. High Energy Phys. {\bf 07}, 012 (2002); D. Stump et al., J. High Energy Phys. {\bf 10}, 046 (2003).

\bibitem{cdf_stw} T. Aaltonen, et al., (CDF Collaboration), public note 10952.

\bibitem{cdf_zpt} T. Aaltonen, et al., (CDF Collaboration), Phys. Rev. D. {\bf 86}, 052010 (2012).

\bibitem{APV} S. C. Bennett and C. E. Wieman, Phys. Rev. Lett. {\bf 82}, 2484 (1999).

\bibitem{milloer} P. L. Anthony et al. (SLAC E158 Collaboration), Phys. Rev. Lett. {\bf 95}, 081601 (2005).

\bibitem{nutev} G. P. Zeller et al. (NuTeV Collaboration), Phys. Rev. Lett. {\bf 88}, 091802 (2002) [Erratum-ibid. 90, 239902(2003)].

\bibitem{lep} G. Abbiendi et al. (LEP Collaborations ALEPH, DEL-PHI, L3 and OPAL; SLD Collaboration, LEP Elec-troweak Working Group, SLD Electroweak and Heavy Flavor Groups), Phys. Rep. {\bf 427}, 257 (2006).

\bibitem{cdf_stw_old} D. Acosta, et al. (CDF Collaboration), Phys. Rev. D {\bf 71}, 052002 (2005).

\bibitem{dzero_1fb} V. M. Abazov, et al. (D0 Collaboration), Phys. Rev. Lett. {\bf 101}, 191801 (2008).

\bibitem{dzero_5fb} V. M. Abazov, et al. (D0 Collaboration), Phys. Rev. D {\bf 84}, 012007 (2011). 

\bibitem{cms_stw} S. Chatrchyan, et al. (CMS Collaboration), Phys. Rev. D {\bf 84}, 112002 (2011)

\bibitem{cs_frame} J. C. Collins and D. E. Soper, Phys. Rev. D {\bf 16}, 2219 (1977).

\bibitem{z_diffx} E. Mirkes, Nucl. Phys. B {\bf 387}, 3 (1992); E. Mirkes and J. Ohnemus, Phys. Rev. D {\bf 50}, 5692 (1994).
 
\bibitem{cdf_angu} T. Aaltonen, et al. (CDF Collaboration), Phys. Rev. Lett. {\bf 106}, 241801 (2011).

\bibitem{pythia} T. Sjostrand, P. Eden, L. Lonnblad, G. Miu, S. Mrenna, and E. Norrbin, Comput. Phys. Commun. {\bf 135}, 238 (2001).

\bibitem{photos} E. Barberio and Z. Was, Computer Phys. Comm. {\bf 79}, 291 (1994); E. Barberio, B. van Eijk, and Z. Was, ibid. {\bf 66}, 115 (1991)

\bibitem{resbos} G. A. Ladinsky and C.-P. Yuan, Phys. Rev. D {\bf 50}, 4239 (1994); C. Balazs and C.-P. Yuan, ibid. {\bf 56}, 5558 (1997); F. Landry, R. Brock, P. M. Nadolsky, and C.-P. Yuan, ibid. {\bf 67}, 073016 (2003); A. Konychev and P. Nadolsky, Phys. Lett. B {\bf 633}, 710 (2006).

\bibitem{wmass_com} J. Beringer et al. (Particle Data Group), Phys. Rev. D {\bf 86}, 010001 (2012).

\bibitem{drellyan} S. D. Drell and T.-M. Yan, Phys. Rev. Lett. {\bf 25}, 316 (1970).

\bibitem{fewz2} K. Melnikov and F. Petriello, Phys. Rev. D {\bf 74}, 114017 (2006).

\bibitem{dzero_wjet} V. M. Abazov, et al. (D0 Collaboration), arXiv:1302.6508 [hep-ex]


\bibitem{cdf_wb} T. Aaltonen, et al. (CDF Collaboration), Phys. Rev. Lett. {\bf 104}, 131801 (2010).

\bibitem{dzero_wb} V. M. Abazov, et al. (D0 Collaboration), Physics Letters B {\bf 718} (2013)

\bibitem{mcfm} J. M. Campbell and R. K. Ellis, Phys. Rev. D {\bf 60}, 113006 (1999); ibid. {\bf 62}, 114012 (2000); ibid. {\bf 65}, 113007 (2002).

\bibitem{sherpa} T. Gleisberg et al., J. High Energy Phys. {\bf 02}, 007 (2009).

\bibitem{madgraph} J. Alwall, et al., J. High Energy Phys. {\bf 1106}, 128, 2011.

\bibitem{cdf_wc} T. Aaltonen, et al. (CDF Collaboration),  Phys. Rev. Lett. {\bf 110}, 071801 (2013).

\bibitem{cdf_zjets} T. Aaltonen, et al., (CDF Collaboration), public note 10216.

\bibitem{cdf_zb} T. Aaltonen, et al., (CDF Collaboration), public note 10594.

\bibitem{dzero_zb} V. M. Abazov, et al. (D0 Collaboration), arXiv:1301.2233 [hep-ex]

\end{thebibliography}
\end{document}